\documentclass[%
 aip,
 amsmath,amssymb,
 reprint,%
]{revtex4-1}

\usepackage{graphicx}
\usepackage{dcolumn}
\usepackage{bm}

\usepackage[utf8]{inputenc}
\usepackage[T1]{fontenc}
\usepackage{mathptmx}
\usepackage{etoolbox}
\usepackage{xcolor}
\usepackage{gensymb}
\usepackage{makecell}
\usepackage{amsmath}
\newcommand\numberthis{\addtocounter{equation}{1}\tag{\theequation}}

\makeatletter
\def\@email#1#2{%
 \endgroup
 \patchcmd{\titleblock@produce}
  {\frontmatter@RRAPformat}
  {\frontmatter@RRAPformat{\produce@RRAP{*#1\href{mailto:#2}{#2}}}\frontmatter@RRAPformat}
  {}{}
}%
\makeatother
\begin{document}

\preprint{AIP/123-QED}

\title[Heterogeneous integration of silicon nitride and amorphous silicon carbide photonics]{Heterogeneous integration of silicon nitride and amorphous silicon carbide photonics}
\author{Zizheng Li}
\email{z.l.li-1@tudelft.nl}
\altaffiliation[]{These authors contributed equally to this work.}

\author{Bruno Lopez-Rodriguez}%
\altaffiliation[]{These authors contributed equally to this work.}
\affiliation{ 
Department of Imaging Physics (ImPhys), Faculty of Applied Sciences, Delft University of Technology, Delft 2628 CJ, The Netherlands
}

\author{Naresh Sharma}
\affiliation{ 
Department of Electronics Engineering, Indian Institute of Technology (Indian School of Mines), Dhanbad, Jharkhand, India
}

\author{Roald van der Kolk}
\affiliation{ 
Department of Imaging Physics (ImPhys), Faculty of Applied Sciences, Delft University of Technology, Delft 2628 CJ, The Netherlands
}
\author{Thomas Scholte}
\affiliation{ 
Department of Imaging Physics (ImPhys), Faculty of Applied Sciences, Delft University of Technology, Delft 2628 CJ, The Netherlands
}

\author{Harmen Smedes}
\affiliation{ 
Department of Quantum Nanoscience, Faculty of Applied Sciences, Delft University of Technology, Delft 2628 CJ, The Netherlands
}

\author{R.Tufan Erdogan}
\affiliation{ 
Department of Precision and Microsystems Engineering, Delft University of Technology, 2628 CD Delft, The Netherlands
}

\author{Jin Chang}
\affiliation{ 
Department of Quantum Nanoscience, Faculty of Applied Sciences, Delft University of Technology, Delft 2628 CJ, The Netherlands
}

\author{Hugo Voncken}
\affiliation{ 
Department of Imaging Physics (ImPhys), Faculty of Applied Sciences, Delft University of Technology, Delft 2628 CJ, The Netherlands
}

\author{Jun Gao}
\author{Ali W Elshaari}
\affiliation{ 
Department of Applied Physics, KTH Royal Institute of Technology, Albanova University Centre, Stockholm 106 91, Sweden
}
\author{Simon Gröblacher}
\affiliation{ 
Department of Quantum Nanoscience, Faculty of Applied Sciences, Delft University of Technology, Delft 2628 CJ, The Netherlands
}
\author{Iman Esmaeil Zadeh}
\affiliation{ 
Department of Imaging Physics (ImPhys), Faculty of Applied Sciences, Delft University of Technology, Delft 2628 CJ, The Netherlands
}

\date{\today}

\begin{abstract}
Amorphous silicon carbide (a-SiC) has emerged as a compelling candidate for applications in integrated photonics, known for its high refractive index, high optical quality, high thermo-optic coefficient, and strong third-order nonlinearities. Furthermore, a-SiC can be easily deposited via CMOS-compatible chemical vapor deposition (CVD) techniques, allowing for precise thickness control and adjustable material properties on arbitrary substrates. Silicon nitride (SiN) is an industrial well-established and well-matured platform, which exhibits ultra-low propagation loss, but it is suboptimal for high-density reconfigurable photonics due to the large minimum bending radius and constrained tunability. In this work, we monolithically combine a-SiC with SiN photonics, leveraging the merits of both platforms, and achieve the a-SiC/SiN heterogeneous integration with an on-chip interconnection loss of 0.32$\pm$0.10 dB, and integration density increment exceeding 4,444-fold. By implementing active devices on a-SiC, we achieve 27 times higher thermo-optic tuning efficiency, with respect to the SiN photonic platform. In addition, the a-SiC/SiN platform gives the flexibility to choose the optimal fiber-to-chip coupling strategy depending on the interfacing platform, with efficient side-coupling on SiN and grating-coupling on a-SiC platform. The proposed a-SiC/SiN photonic platform can foster versatile applications in programmable and quantum photonics, nonlinear optics, and beyond.  
\end{abstract}

\maketitle

\section{\label{sec:level1}Introduction}




Photonic integrated circuits (PICs), with the merits including miniaturization, robustness, and high-volume manufacturability, have been one of the cornerstones for applications in data processing, optical communication, and hold potentials in medical imaging, sensing, and quantum computing.\cite{shekhar2024roadmapping,lu2024empowering,perez2020principles,yao2023integrated,elshaari2020hybrid,pelucchi2022potential} 
Various material platforms have been studied and utilized as the foundations for constructing photonic integrated circuits. Among all material platforms which are proposed and researched, silicon (Si) and silicon nitride (SiN) photonics have been the most matured PIC platforms available for mass production in the industry. 

Leveraging the micro-electronics fabrication techniques, silicon (Si) photonics has flourished over the past decades.\cite{jalali2006silicon} Si's high refractive index and large thermo-optics coefficient enable high integration density and low-power tunable photonic devices.\cite{komma2012thermo,li1980refractive,johnson2022determination} However, the presence of two-photon absorption in silicon hinders silicon photonics from scaling up to very large photonic networks, so that prevents the realization of more advanced applications.\cite{bristow2007two,tsang2002optical} Silicon's relatively narrow energy band gap ($\sim$ 1.1 eV) also results in high absorption losses at visible wavelengths, limiting silicon photonics to, mostly, the telecommunication wavelengths.\cite{alex1996temperature,passler1996comparison} Silicon nitride (SiN), on the other hand, provides ultra-low waveguide propagation losses in a wide transparency window covering both visible and telecommunication wavelengths, thereby facilitating versatile photonic devices and applications.\cite{puckett2021422,liu2021high,chauhan2022ultra,corato2024absorption} In addition, SiN offers a relatively large Kerr coefficient.\cite{klanjvsek2001infrared,ikeda2008thermal} These characteristics together make SiN an excellent host for nonlinear optics applications, including, wavelength conversion, entanglement generation, frequency combs, and more.\cite{ikeda2008thermal,liu2021high,kim2017dispersion,frigg2020optical,choi2018broadband,yin2021frequency,zhu2024enhanced,xiong2015compact,duan2024visible,lu2019chip} To realize the ultra-low loss SiN waveguides, high-aspect-ratio waveguides (width:height > 10:1) have been proposed and successfully fabricated using either the subtractive or Damascene processes, where the optical modes are weakly confined around the waveguide slab to exploit the low-absorption advantage of silicon dioxide (SiO$_2$) cladding. To fabricate such ultra-low loss SiN waveguides, high-temperature ($\geq$ 1200 $\celsius$) annealing and chemical-mechanical polishing (CMP) are essential to effectively minimize the loss.\cite{heck2014ultra,puckett2021422,liu2021high,bauters2011ultra,huffman2018integrated} Owing to SiN's relatively low refractive index, large minimal bending radius (on the millimeter to sub-centimeter scale) is necessary, which compromises the integration density. In addition, to prevent metal absorption loss the high-aspect ratio SiN platform requires a sufficiently thick cladding layer on top of the waveguides (6-15 $\mu$m) that separates the SiN waveguides from the micro-heaters when employing thermo-optic tuning, this hampers SiN photonics from pursuing efficient tunability and poses challenges in thermal crosstalk control. Combined with the low intrinsic thermo-optic coefficient of SiN, it is challenging to achieve efficient thermo-optic tuning on high-aspect-ratio SiN planform.\cite{kleiner2002thermal,elshaari2016thermo,arbabi2013measurements} Therefore, despite the prevalence and maturity of Si and SiN photonics, their performance ceiling in scaling-up and effective tuning has driven the demand of investigating new photonic platforms.

Amorphous silicon carbide (a-SiC) is a promising photonic platform that emerged recently, appealing for its properties involving large and tunable band gap ($\sim$ 2.5 eV), relatively high and tunable refractive index ($\sim$ 2.4 - 2.9 at 1550 nm), low loss (0.78 dB/cm), large thermo-optic coefficient ($1.12\times10^{-4}/\celsius$), and large Kerr-nonlinear coefficient ($6.7\times10^{-18}$ $\text{m}^2/\text{W}$).\cite{lopez2023high,lopez2025magic,xing2019cmos,xing2020high,lu2024strong,chang2022demonstration,lu2014optical,xing2019cmos} However, despite the advancements achieved in the recent literature, a-SiC exhibits higher optical loss compared to SiN. Heterogeneous integration of a-SiC on lithium niobate via low-temperature CVD deposition has been recently realized, showing that a-SiC's low-temperature fabrication techniques are compatible with the different substrate, with precise thickness control, and excellent surface roughness without polishing.\cite{li2025heterogeneous} 

In various PIC platforms, fiber-to-chip interfacing has remained a persistent topic in integrated photonics research. Side-coupling (or edge-coupling), which is known for its high coupling efficiency and large bandwidth, has been extensively studied by researchers over the decades since the origin of integrated photonics.\cite{alder2000high,marchetti2019coupling,hauffe2001methods,fernandez2019low,zhang2023efficient,son2018high,mu2020edge} The primary challenge of side-coupling is the substantial mode mismatch between the modes in optical fibers and photonic waveguides. Delicate design and fabrication are needed to realize the adiabatic transition from the strongly confined waveguide modes into weakly confined modes that match that of the fiber modes.  
Grating-coupling is another widely adopted fiber-to-chip coupling strategy, which obviates the need for chip-facet exposure and additional waveguide routing, thereby simplifies the fabrication process. It also outperforms in alignment tolerance, compared to edge-coupling. These facts make grating-coupling inherently compatible with wafer-scale production and characterization. Designing high-coupling-efficiency grating couplers requires an optimal grating period and favors a large film thickness and high refractive index contrast.\cite{son2018high,cheng2020grating,lin2023ultra,korvcek2023low,kohli2023c,chmielak2022high} With above arguments, high-aspect-ratio SiN waveguide, the record-low propagation loss holder, is unsuitable for achieving high-efficiency grating-coupling. On the other hand, in high refractive index material platforms, such as a-SiC, it is challenging to realize efficient edge-coupling. This is not the case when the roles are reversed, each platform becomes well-suited for the other's task. For instance, the high-aspect-ratio SiN's weak confinement leads to a large mode area, and makes the mode matching easier, while a-SiC's high refractive index offers more design flexibility for grating couplers. The proposed a-SiC/SiN heterogeneous integration provides a unified solution for fiber-to-chip interfacing, enabling both grating-coupling and edge-coupling to be employed optimally. 

Thermo-optic tuning is widely exploited to actively control the on-chip photonic devices in large-scale PICs for photonic computing, quantum computing.\cite{xie2024towards,xu2024large,bao2023very,wang2025asymmetrical,qiang2018large,ahmed2025universal,xu2022self,yao2023integrated,bogaerts2020programmable,luo2023recent,pelucchi2022potential} Thermal tuning efficiency and power consumption are recognized as the pivotal performance benchmarks of a PIC, meanwhile the scalability is also highly valued, which is mainly limited by waveguide losses. The a-SiC/SiN heterogeneous photonic platform leverages the high thermal optical coefficient of amorphous silicon carbide, empowering the ultra-low loss high-aspect-ratio SiN photonics with efficient thermo-optic tunability, together they present an effective and technically feasible approach towards a scalable, densely integrated, ultra-low loss, and highly tunable photonic platform. 

In this paper, we propose a heterogeneous a‑SiC/SiN photonic integrated circuit that combines the high-aspect-ratio SiN waveguides with efficiently thermo-optic tunable and high integration density a-SiC waveguides. By showcasing the key building blocks, including low-loss a‑SiC/SiN interconnection couplers and fiber-to-chip grating couplers, we outline potential directions for future applications enabled by this platform, taking advantages of its ultra-low loss, wide transparency window, high integration density, efficient thermo-optic tunability, and strong third-order nonlinearity properties. 

\begin{figure*}[ht!]
\centering\includegraphics[width=14cm]{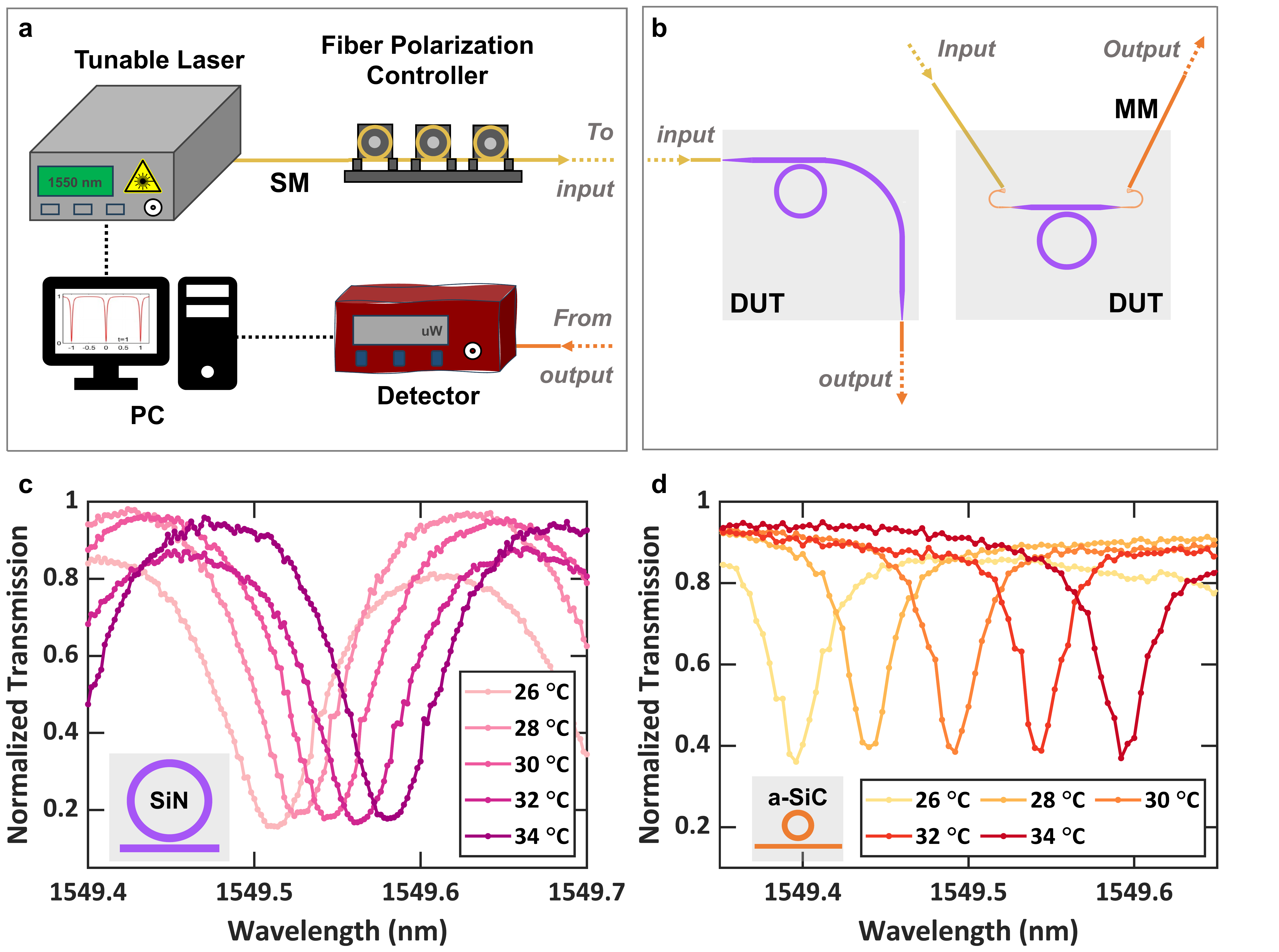}
\caption{(a) The characterization setups used to measure on-chip devices. (b) Two kinds of fiber-to-chip accesses employed in measurements. Thermo-optic characterization of the a-SiC/SiN platform: the resonance shifts observed when sweeping the temperature from 26 \celsius\ to 34 \celsius\ on the (c) SiN ring resonator and (d) a-SiC ring resonator.} 
\end{figure*}

\section{Photonic devices and performance characterization}
Two different setups, one for side-coupling and the other for grating-coupling, are employed to characterize the performance metrics of a-SiC/SiN platform, including thermal optics response, a-SiC/SiN interconnection coupling, edge-coupling and grating-coupling. Fig.1(a) and (b), and Fig.S1 in Supplementary Material show the corresponding setups. 
The characterization is performed with tunable lasers (Photonetics TUNICS-PRI 3642 HE 15 and Keysight 81940A). The light is sent to the photonic chip via a single-mode fiber, passing through a paddle polarization controller (FPC560). The output light is collected by a multi-mode fiber and delivered to an optical powermeter (Newport 818-NR) or photoreceiver (Newport 1811 New Focus). For grating-coupling, the fibers are positioned at an angle of 10 degrees relative to the normal of the sample. For side-coupling, we use translation stages to align optical fibers to the exposed waveguide cross-sections. To measure the thermo-optics response, the sample is mounted and pasted on the sample stage, using thermally conductive silver paste. Heat is applied to the sample stage by a proportion-integration-differentiation temperature controller to globally adjusting the temperature of the whole chip. A temperature sensor in the stage provides feedback to the temperature controller to maintain a stable temperature. To characterize the interconnection coupling efficiency, we designed and fabricated the devices shown in the inset of Fig.2(d). The interconnection devices consist of two grating couplers as input and output, and two interconnection couplers connecting the a-SiC and SiN waveguides, while the reference devices (Ref.) are made of two grating couplers connected by a straight a-SiC waveguide. The definitions of the characterized performance metrics are discussed in detail in Method.

\subsection{Thermal optics}
On-chip tunability after fabrication is vital for all applications that require programmability and reconfigurability. As a key feature of the proposed a-SiC/SiN platform, the thermo-optic tunability is characterized utilizing the on-chip ring resonators. We apply global heating to the a-SiC/SiN sample and monitor the resonance wavelength shift with respect to the temperature variation, using the characterization setup shown in Fig.1(a) and (b). On the same sample, a-SiC microrings with a bending radius of 120 $\mu$m and SiN microrings with a bending radius of 1100 $\mu$m are fabricated and measured. The a-SiC rings demonstrate a 25.0 pm/$^\circ$C shift over the temperature change from 26 $^\circ$C to 34 $^\circ$C, whereas the measured SiN ring spectra show a 8.6 pm/$^\circ$C temperature shift over the same range, as illustrated in Fig.1(c) and (d). Considering that the SiN ring resonators possess a much larger bending radius, hence a much larger heating area, we normalize the thermo-optic shift to the unit ring-waveguide length, and taking it as the metic to evaluate thermo-optic tunability. According to the measured results, a-SiC ring shows a thermo-optic tunability of $33.2\times10^{-3}$ pm/$\celsius$/$\mu$m, while SiN ring shows $1.2\times10^{-3}$ pm/$\celsius$/$\mu$m. It is concluded that a-SiC has 27 times stronger thermo-optic tunability compared to SiN. A thermo-optic coefficient of $5.1 \times 10^{-5}/\celsius$ for a-SiC, and $3.6 \times 10^{-5}/\celsius$ for SiN, are extracted (details in Method), showing good agreement with the recent works \cite{lopez2025magic,chang2023high,chang2023high,johnson2022determination}. The heterogeneous integration of a-SiC and SiN opens the way for ultra-low-loss SiN photonics to attain efficient tunability while providing highly programmable a-SiC photonics with low-loss delay lines, among other possible advanced functionalities.

\subsection{a-SiC/SiN interconnection couplers}


\begin{figure*}[ht!]
\centering\includegraphics[width=13cm]{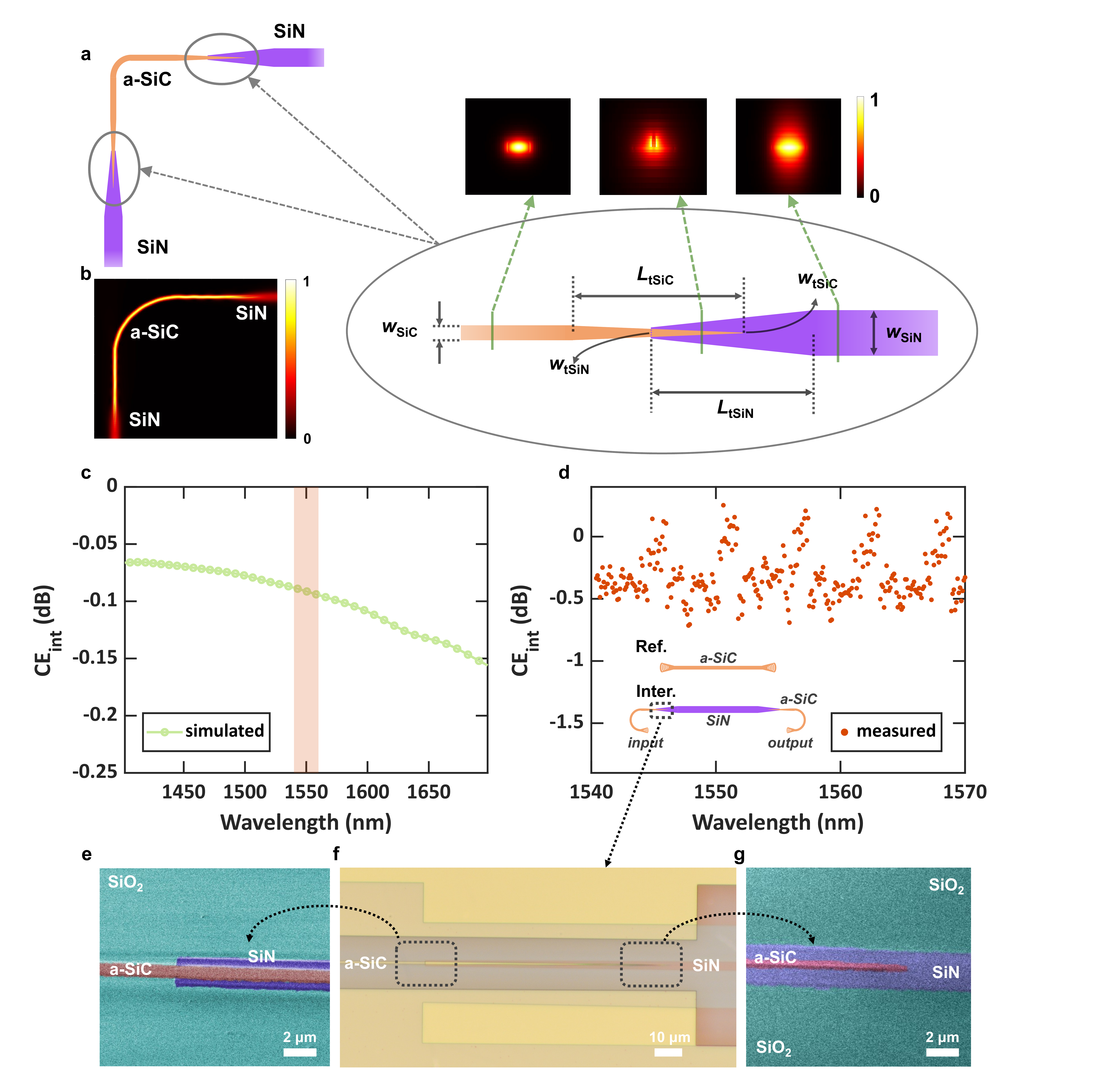}

\caption{(a) A schematic illustration of the a-SiC/SiN interconnection coupler, with a zoomed-in figure showing the interconnection coupling region. Three insets are simulated mode profiles in the a-SiC waveguide, at the coupling region, and in the SiN waveguide, respectively. (b) Simulated electric field intensity distribution of the light propagating between SiN and a-SiC layers, through the a-SiC 90-degree bending and interconnection couplers. (c) Simulated interconnection coupling efficiency, with a light-orange strip showing the wavelength range of experimental measurements. (d) Measured interconnection coupling efficiency. The insets depict the fabricated reference device (Ref.) and interconnection coupler (Inter.). (e) Scanning electron microscopy image of the SiN tip in the interconnection coupler. (f) Optical microscopy image of the interconnection coupler. (g) Scanning electron microscopy image of the a-SiC tip in the interconnection coupler.}
\end{figure*}

The a-SiC/SiN interconnection coupler that enables bidirectional light coupling between a-SiC and SiN is shown in Fig.2(a). It adiabatically converts the mode propagating in a-SiC waveguides into the mode supported by SiN waveguides, and vice versa. The design principle of such devices is described in our previous works.\cite{li2023heterogeneous,sharma2024design} The ultra-low loss SiN platform relies on the high aspect ratio waveguides, where a greater aspect ratio is proven to provide lower loss.\cite{huffman2018integrated} At telecom wavelengths, the dimensions of single mode waveguides are 100 nm $\times$ 3 $\mu$m. In contrast, the a-SiC single mode waveguide has a size of 260 nm $\times$ 800 nm.\cite{lopez2023high} The refractive index contrast between a-SiC and SiN, reaching $\Delta n =$0.6 around 1550 nm, along with the 160 nm thickness difference of a-SiC/SiN waveguides, results in a significant mode mismatch. To overcome this mismatch and achieve phase matching, both a-SiC and SiN waveguides are adiabatically tapered down to the final tip widths. The widths of the a-SiC and SiN waveguides are represented as $w_\text{SiC}$ and $w_\text{SiN}$, respectively, in Fig.2(a) inset. As shown in the zoomed-in illustration, the a-SiC and SiN waveguides are tapered towards the center of the device, with their widths decreasing to the final tip widths of $w_\text{tSiC}$ and $w_\text{tSiN}$, respectively. 

The tapering lengths are denoted as $L_\text{tSiC}$ and $L_\text{tSiN}$, respectively. We use finite-difference-eigenmode method to calculate the optical mode profiles, and finite-difference time-domain (FDTD) method to simulate light propagation in the a-SiC/SiN interconnection coupler. The mode profiles of the light propagating in the a-SiC waveguide, the a-SiC/SiN coupling region, and the SiN waveguide are shown in the three insets at the top right, from left to right, respectively. We introduce a 90-degree turn to determine and verify the minimal bending radius of a-SiC, and found a minimal bending radius smaller than 10 $\mu$m. The electric-field intensity distribution in the device, in which light launched into a SiN waveguide is coupled to an a-SiC waveguide, guided through a 90° bend, and subsequently coupled back to another SiN waveguide, is shown in Fig.2(b). The coupling efficiency of the a-SiC/SiN interconnection coupler, $CE_\text{int}$, is simulated by FDTD method (Ansys Lumerical) as shown in Fig.2(c), with a light orange strip highlighting the wavelength range that is experimentally investigated. The standard single-mode a-SiC and SiN waveguides have widths of 0.8 $\mu$m and 3 $\mu$m, respectively. The optimal taper tip widths are $w_\text{tSiC}=50$ nm and $w_\text{SiN}=2$ $\mu$m, according to simulations. The results exhibit that the interconnection loss is less than 0.1 dB per connection at 1550 nm.

Experimentally, we characterized the bending losses of a-SiC waveguides by measuring the intrinsic quality factors of the ring resonators with different bending radii, and discovered the minimal bending radius can be smaller than 15 $\mu$m (Supplementary Material). We believe that the minimal bending radius can be reduced to below 10 $\mu$m by optimizing the single mode waveguide geometry. As comparison, the minimal bending radius for the 100 nm thick high-aspect ratio SiN waveguide is larger than 1 mm\cite{huffman2018integrated}. This indicates that combining a-SiC/SiN photonics on the same chip enables compact photonic routing and smaller device footprint, and hence significantly increases the integration density, which is essential for scalable PIC design. Considering the photonic devices (for example ring resonators, spirals, and routing bends) scale with the square of the minimal bending radius $R^2$, the integration density $D$ can be evaluated as $D \propto R^2$. The experimental results reveal an 4,444-fold integration density difference between a-SiC/SiN photonics and SiN photonics. Regarding the a-SiC/SiN interconnection coupling, as shown in Fig.2(d), a coupling loss of 0.32 $\pm$ 0.10 dB per a-SiC/SiN connection is measured from the devices with the optimal parameters suggested by the simulation results. The results are obtained via normalizing the averaged output power from four identical interconnection couplers to the output power of the reference devices (detailed in Methods). Fabry-Perot (F-P) oscillations are observed superimposed on the signal. Based on the period, F-P oscillations are considered to arise from the reflections between fiber and grating coupler. Fig.2(f) shows the optical microscope image of the fabricated a-SiC/SiN interconnection coupler, while the a-SiC and SiN taper end tips are inspected using a scanning electron microscope, and shown in Fig.2(e) and (g), respectively.

To further study the fabrication tolerance, we sweep the parameters of a-SiC tip width $w_\text{tSiC}$ from 30 nm to 70 nm, and SiN tip width $w_\text{tSiN}$ from 1.4 $\mu$m to 2.4 $\mu$m. The measurement results are displayed respectively in Fig.S3 in Supplementary Materials. For each of the parametric sweeps, 5 devices are fabricated on the same sample. The sweeping results of $w_\text{tSiC}$ and $w_\text{tSiN}$ show a maximum extra loss of 1.1 dB in $CE_\text{int}$. The change in a-SiC tip width $w_\text{tSiC}$ results in a more significant impact on $CE_\text{int}$. In general, the a-SiC/SiN exhibits good fabrication tolerances. 
The same design protocol can also be applied to the a-SiC/SiN integration platform with other layer thicknesses, for SiN waveguides with different aspect ratios and applications in a variety of wavelengths.

\subsection{Edge couplers}

\begin{figure*}[ht!]
\centering\includegraphics[width=11cm]{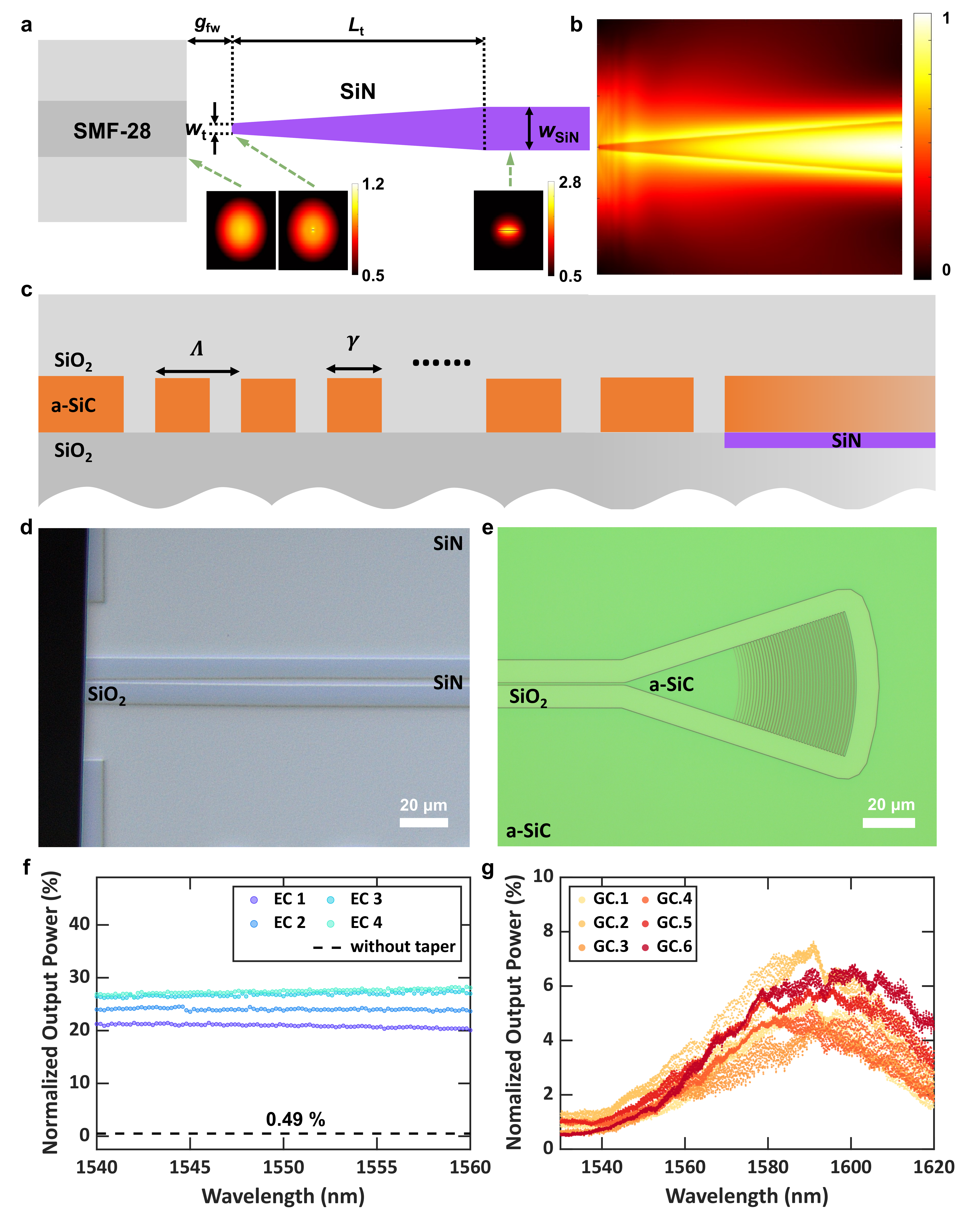}
\caption{(a) Schematic showing the fiber-to-chip coupling using inverted taper edge coupler, with taper end tip width $w_\text{t}$ and taper length $L\text{t}$. $g_\text{fw}$ represents the fiber-to-waveguide gap. (b) Simulated electric field intensity distribution in the edge coupling process. (c) Schematic of the apodization grating coupler, with grating period $\Lambda$ and grating width $\gamma$. (d) An optical microscopy image showing the fabricated edge coupler. (e) Optical microscopy image showing the fabricated grating coupler. (f) The measured power from devices' output, after passing through the input edge coupler (EC), 90 degrees waveguide bend, and the output edge coupler, which is normalized to laser power. On the same chip, four identical devices with the optimized parameters are measured. The dashed line shows the coupling of the standard single-mode waveguides without inverted tapers. (g) The measured output power, after passing through two grating couplers (GCs) connected by a straight waveguide, which is normalized to the laser power. }
\end{figure*}

One of the most commonly used fiber-to-chip coupling method is edge-coupling (or side-coupling), which provides a higher coupling efficiency and wider bandwidth compared to grating couplers. To overcome the prominent challenge of mode mismatch between the single-mode fiber and the optical waveguide, we design and optimize an inverted taper mode converter. At the wavelength of 1550 nm, the mode area in the standard SMF-28 single-mode fiber is roughly 75.1 $\mu$m$^2$, by contrast, the mode area in a-SiC single-mode waveguide is 0.4 $\mu$m$^2$, while in the high aspect ratio SiN waveguide it is 4.1 $\mu$m$^2$. The high width-to-height ratio can be leveraged to further promote the weak-confinement feature by tapering down the waveguide width, thereby enlarging the waveguide mode area and mitigating the mode mismatch between the fiber and waveguide. As shown in Fig.3(a), the SiN waveguide is tapered down to the optimal tip width $w_\text{t}=570$ nm where the best mode matching and the highest mode overlapping are realized, indicating a theoretical coupling efficiency up to 91.9$\%$ per fiber-to-chip interface when fiber-to-waveguide gap is zero ($g_\text{fw}=0$). The simulated power distribution of the fiber-to-chip coupling process is shown in Fig.3(b).

To experimentally investigate the performance of the designed edge couplers, we fabricate four identical devices, each consisting of two edge couplers connected by a 90-degree waveguide bend. The sample is cleaved twice perpendicularly to expose the cross-sections of the edge couplers. The measurement is done using a single-mode fiber as input injecting light to one of the edge couplers, and a multi-mode fiber collecting from the output, as shown in Fig.1(a) and (b). The measured output power, normalized to the laser, is presented in Fig.3(f). Two devices consisting of cleaved standard SiN single-mode waveguides connected by a 90-degree bend, without tapered edge couplers, are taken as a reference. The measured results are averaged and represented by the black dashed line in Fig.3(f). The measured results show that up to 27.45 $\%$ power of the input can be collected at the output, after experiencing two fiber-to-chip interfaces. In comparison, without the taper only 0.49$\%$ power is collected at the output, indicating a 55-fold improvement in coupling efficiency with the inverted taper structure. The details of the characterization and analysis are provided in Method.
More microscopy pictures demonstrating the sample edges after cleaving are included in Supplementary Material, along with a simulation analysis on how the fiber-to-chip air gap and fiber misalignment are influencing the coupling efficiency. 

\subsection{Grating couplers}
  The other commonly used component to realize fiber-to-chip coupling is grating couplers, which excel in alignment tolerance, and convenient fiber-to-chip access, and are well-suited for multi-input and multi-output reconfigurable photonics. The automatic wafer-scale characterization of a large number of devices also benefits from this. On the a-SiC/SiN platform, we utilize a standard a-SiC grating coupler to couple light from SMF-28 single-mode fiber to the waveguides, and a multi-mode fiber is employed to collect the output from another grating coupler. Fig.3(c) depicts the apodized grating coupler, with parameters of grating period $\Lambda$ and grating line width $\gamma$. We fabricated six grating coupling devices (Ref. configuration as shown in Fig.2(d)) with the same period $\Lambda=1.0$ $\mu$m, and grating line width apodized from $\gamma_0 = 600$ nm to $\gamma_{20} = 940$ nm for GC.1 to 3, and $\gamma_0 = 720$ nm to $\gamma_{20} = 960$ nm for GC.4 to 6. The six devices are measured and the results are shown in Fig.3(g). A maximum normalized output power of 7.7$\%$ is achieved at 1590 nm, while values exceeding 2.4$\%$ are measured around 1550 nm. The detailed definition of the coupling efficiency can be found in Method. F-P oscillations with a period similar to that in Fig.2(d) are observed, which are attributed to the reflections between the fiber and the grating coupler.

\section{Method}

The a-SiC/SiN interconnection coupler efficiency is obtained based on the averaged measurement results that were obtained from multiple identical devices. Assuming that the two interconnection couplers within the same device hold identical optical responses, the interconnection coupling efficiency $CE_\text{int}$ can be defined in the following equation: 
\begin{eqnarray}
P_{\text{out}} = P_{\text{in}} \cdot CE_{\text{gc-in}} \cdot \alpha_{\text{a-SiC}} \cdot \alpha_{\text{SiN}} \cdot CE^2_{\text{int}} \cdot CE_{\text{gc-out}}\text{,}
\label{eq:one}
\end{eqnarray}
while the reference device can be expressed by: 
\begin{eqnarray}
P_{\text{out}}^{\prime} = P_{\text{in}} \cdot CE_{\text{gc-in}} \cdot \alpha^{\prime}_{\text{a-SiC}} \cdot CE_{\text{gc-out}}.
\label{eq:two}
\end{eqnarray}
In Eq.(1) and (2), $P_{\text{in}}$, $P_{\text{in}}^{\prime}$, $P_{\text{out}}$, and $P_{\text{out}}^{\prime}$ denote the input and output power in the optical fibers, respectively. $\alpha_{\text{a-SiC}}$ and $\alpha_{\text{SiN}}$ represent the losses in the a-SiC and SiN waveguides, respectively in the interconnection coupler. 
$CE_{\text{gc-in}}$ and $CE_{\text{gc-out}}$ are the coupling efficiencies of the input and output a-SiC grating couplers, respectively. 
Assuming that on the same sample the grating couplers have the same response, the a-SiC/SiN interconnection coupler coupling efficiency can be written as the following equation. 
\begin{eqnarray}
CE_{\text{int}} = \sqrt{ \frac{P_{\text{out}}}{P_{\text{in}} \cdot \alpha_{\text{a-SiC}} \cdot \alpha_{\text{SiN}} \cdot CE_{\text{gc-in}} \cdot CE_{\text{gc-out}}  } } \nonumber\\
= \sqrt{ \frac{P_{\text{out}} \cdot \alpha^{\prime}_{\text{a-SiC}}} {\alpha_{\text{a-SiC}} \cdot \alpha_{\text{SiN}} \cdot P^{\prime}_{out}  } }
\label{eq:three}
\end{eqnarray}

Regarding grating-coupling, devices with two grating couplers connected by an a-SiC waveguide are fabricated, where the light coupling process is expressed as:
\begin{eqnarray}
P_\text{gc-out} = P_\text{in} \cdot CE_\text{gc-in} \cdot \alpha_\text{gc-SiC} \cdot CE_\text{gc-out}\text{,}
\label{eq:six}
\end{eqnarray}
where $\alpha_\text{gc-SiC}$ is the loss generated by the connecting a-SiC waveguide. The normalized output power is used to quantify the grating-coupling efficiency:
\begin{eqnarray}
P_\text{gc-norm} = \frac{P_\text{gc-out}}{P_\text{in} \cdot \alpha_\text{gc-SiC}}\text{.}
\label{eq:seven}
\end{eqnarray}

For edge coupling efficiency measurement, devices that consist of a $L$ long SiN waveguide and two edge couplers are fabricated. The light coupling can be expressed as:
\begin{eqnarray}
P_\text{sc-out} = P_\text{in} \cdot CE_\text{sc-in} \cdot \alpha_\text{sc-SiN} \cdot CE_\text{sc-out}\text{,}
\label{eq:four}
\end{eqnarray}
where $\alpha_\text{sc-SiN}$ is the loss generated by the connecting SiN waveguide. $CE_\text{sc-in}$ and $CE_\text{sc-out}$ represent the input and output side-coupling efficiencies, respectively. The normalized output power used for quantifying the side-coupling efficiency is defined as:
\begin{eqnarray}
P_\text{sc-norm} = \frac{P_\text{sc-out}}{P_\text{in} \cdot \alpha_\text{sc-SiN}}\text{.}
\label{eq:five}
\end{eqnarray}

The waveguide losses are characterized by the intrinsic quality factor of the ring resonators fabricated on the a-SiC layer and SiN layer, respectively, on the same sample. We translate the intrinsic quality factor $Q_\text{i}$ into propagation loss $\alpha$, in the unit of dB/cm, as expressed below: 
\begin{eqnarray}
Q_{\text{i}} = \frac{2 Q_{\text{L}}}  {1 + \sqrt{T}} = \frac{2 \lambda_0}  { FWHM (1 + \sqrt{T}) } 
\label{eq:eight}
\end{eqnarray}

\begin{eqnarray}
\alpha \, \text{(dB/m)} = 4.3429 \cdot \alpha \text{(m} ^{-1}) = 4.3429 \cdot (\frac{2\pi n_g}{Q_i\lambda_0})
\label{eq:nine}
\end{eqnarray}
 where $\lambda_0$ denotes the center wavelength. The full-width half-maximum (FWHM) of the resonance dips and the on resonance transmission $T$ are measured experimentally. $n_\text{g}$ is the group index of the propagation mode, obtained from the free spectral range (FSR) of the ring resonator:
 \begin{eqnarray}
n_\text{g} = \frac{\lambda^2}{2\pi R \cdot FSR}
\label{eq:ten}
\end{eqnarray}

To characterize the thermo-optic coefficients of a-SiC and SiN photonics, we use the methods introduced in the literature.\cite{elshaari2016thermo,lopez2023high,chang2023high,qiu2015athermal} Eq.(11) below relates the effective index change along temperature change with the material thermo-optic coefficients by: 
\begin{align*}
\frac{\text{d}n_{\text{eff}}} {\text{d}T} &= \Gamma_{\text{SiO}_2} \frac{\text{d}n_{\text{SiO}_2}} {\text{d}T} + 
\Gamma_{\text{a-SiC}} \frac{\text{d}n_{\text{a-SiC}}} {\text{d}T}
\\
\frac{\text{d}n_{\text{eff}}} {\text{d}T} &= \Gamma_{\text{SiO}_2} \frac{\text{d}n_{\text{SiO}_2}} {\text{d}T} + 
\Gamma_{\text{SiN}} \frac{\text{d}n_{\text{SiN}}} {\text{d}T}
\numberthis \label{eq:eleven}
\end{align*}
where $\Gamma$ denotes the mode overlap integral in the corresponding material, which is obtained from finite difference eigenmode simulations,and the thermo-optic coefficient of SiO$_2$ is $0.96 \times 10^{-6}/\celsius$ as determined in the literature \cite{elshaari2016thermo}. The effective index change of the fundamental waveguide mode can be obtained from the measured resonance wavelength shift, given by equation:
\begin{eqnarray}
\frac{\text{d}\lambda}{\text{d}T} = (n_\text{eff}\cdot \alpha_{\text{SiO}_2} + \frac{\text{d}n_\text{eff}}{\text{d}T})\cdot \frac{\lambda}{n_\text{g}}
\label{eq:twelve}
\end{eqnarray}
where $\alpha_{\text{SiO}_2} = 2.6\times10^{-6}/\celsius$ is the thermal expansion coefficient of the silicon dioxide.

\section{Prospects of a-SiC/SiN platform for future applications}
\begin{figure*}[ht!]
\centering\includegraphics[width=15cm]{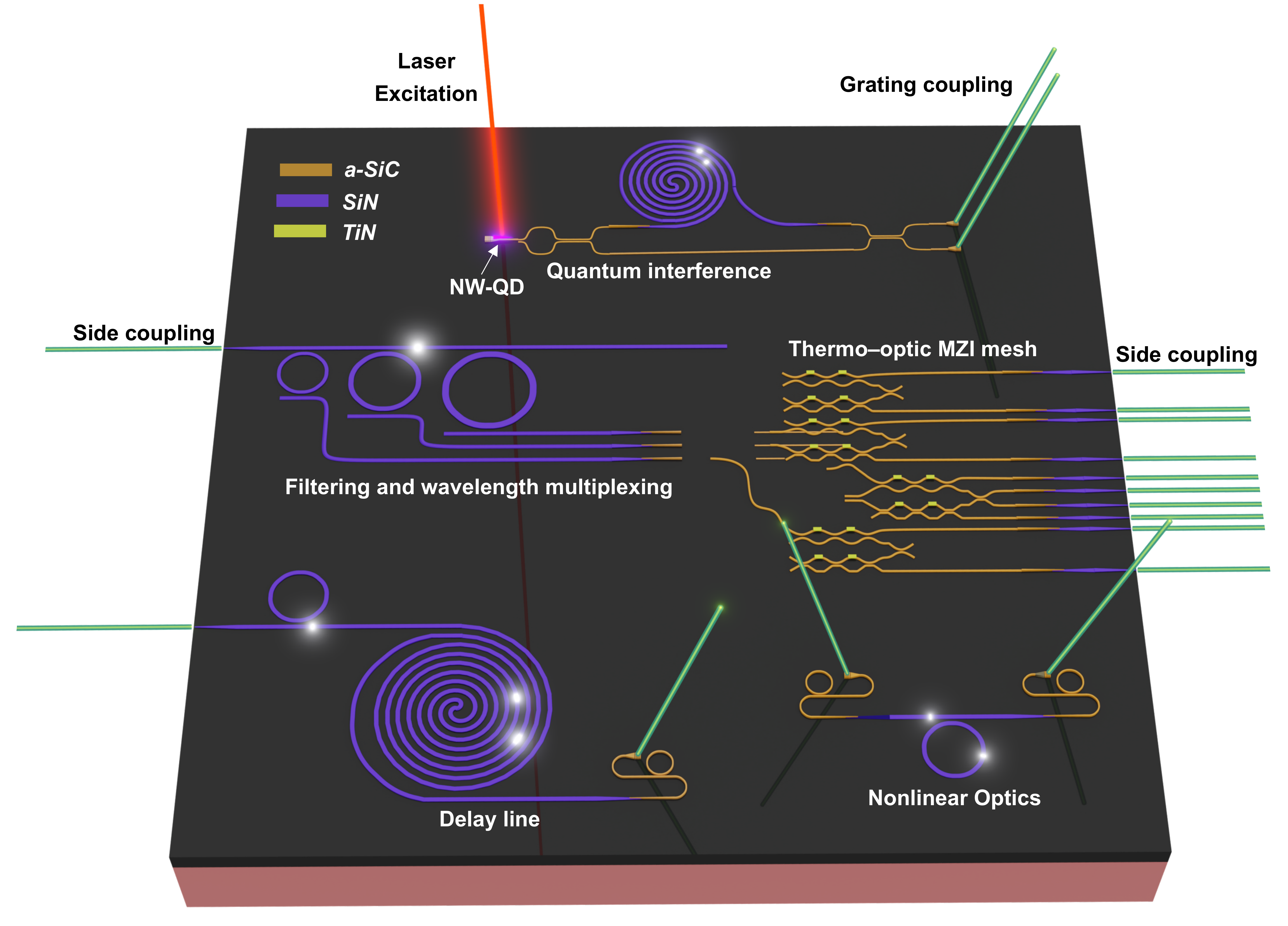}
\caption{An illustration of the a-SiC/SiN heterogeneous photonic integrated circuit with the functional devices and potential applications. Glowing white dots illustrate photons propagating in the waveguides. Optical fibers are represented using green color, and the free-space laser beam is depicted with red color.} 
\end{figure*}

Examples of the future applications of the proposed heterogeneous a-SiC/SiN PIC are depicted in Fig.4, where we mark a-SiC in orange, SiN in dark blue, titanium nitride (TiN) in yellow, SiO$_2$ in black, and Si substrate in red. 
Optical fibers (in green) are positioned either toward the exposed sample facets for edge-coupling, or angled at 20$\degree$ from above for grating-coupling. Employing the pick-and-place technique, we foresee the immediate implementation of deterministic integration of nanowire quantum dots (NW-QDs) as single photon emitters, and subsequently quantum optics experiments on the a-SiC/SiN platform\cite{chanana2022ultra,dusanowski2023chip,zadeh2016deterministic,chang2023nanowire,mnaymneh2020chip}. Devices delay lines with tens of nanoseconds delayed time
with the distinctive advantage of combining high density and high tunability a-SiC photonics with ultra-low-loss SiN photonics, a-SiC/SiN photonic platform can emerge as a compelling choice for scalable and reconfigurable photonics.

TABLE I lists the comparison of the key performance metrics of different tunable PIC platforms, including integration density ($\rho=1/R_\text{min}^2$), waveguide loss, thermo-optic coefficient, power consumption for $\pi$ phase shift, and fabrication cost per minimal device area. Here, we use the square of the minimal bending radius ($R_\text{min}^2$) to represent the minimal device area. In the table, we focus on single-mode waveguides at telecom wavelengths, therefore, strongly confined (thick-film) platforms are excluded from comparison. More detailed sources of these estimations are provided in Supplementary Material. As presented in the table, Si PIC offers the highest integration density and thermo-optic tunability, albeit at the cost of high waveguide loss. The SiN PIC platforms, namely the moderate confinement and high-aspect-ratio SiN, exhibit low waveguide loss, but suffer from high fabrication expense and thermal tuning power consumption. In comparison, the a-SiC/SiN PIC possesses high integration density and efficient thermal tunability that are comparable to Si PIC, while it retains the low-loss characteristic of high-aspect-ratio SiN PIC. Owing to the substantial difference in integration densities, the heterogeneous a-SiC/SiN platform also shows a notable advantage in fabrication price per minimal device area ($R_\text{min}^2$). 

\begin{table*}
\caption{\label{tab:table1} Comparison of key performance metrics between a large-scale a-SiC/SiN photonic network and other material platforms.}

\begin{ruledtabular}
\begin{tabular}{ccccccc}
  
  \footnote{Standard 220 nm silicon on insulator platform, SiN moderate confinement (SiN-M) platform with a waveguide thickness of 300$\sim$400 nm, and SiN high-aspect-ratio (SiN-H) platform with waveguide thickness $\leq$ 100 nm, at the wavelength of 1550 nm. The a-SiC/SiN platform proposed in this work combines 260 nm a-SiC photonics with 100 nm high-aspect-ratio SiN} Platform &
  \makecell{\footnote{Defined as $\rho=1/R_\text{min}^2$, where $R_\text{min}$ denotes the minimal bending radius of the platform.}Integration density\\ ($/$mm$^{2}$)} & 
  \makecell{Waveguide loss \\ (dB/cm) } &
  \makecell{Thermo-optic\\coefficient (/$\celsius$)} &
  \makecell{Power consumption per\\$\pi$ phase shift (mW)} &
  \makecell{\footnote{The method used to estimate prices is detailed in Supplementary Material. }Fabrication price per\\minimal device area (\$$\cdot \text{mm}^2$)}&
  Ref. \\
 
 \hline

 a-SiC/SiN  & 4,444 & 0.001 $\sim$ 0.78 & $5.1\times10^{-5}$ & 35 & 0.229 & This work,\citenum{lopez2023high,puckett2021422}

\\Si & 40,000 & 1.5$\sim$4 & $1.8\times10^{-4}$ & 20 & 0.013 & \citenum{liu2022thermo,parra2024silicon,xie2020thermally}

\\SiN (M) & 100 & 0.15$\sim$1.5 & $2.5\times10^{-5}$ & 88 & 5.080 & \citenum{sacher2015multilayer,li2023process,bonneville2021low,frigg2019low,nejadriahi2021efficient,tyler2019sin}

\\SiN (H) & 1 & 0.001$\sim$0.1 & $2.5\times10^{-5}$ & 300 & 614.644 & \citenum{hai2015thermally,spencer2014integrated,puckett2021422,chauhan2022ultra,chanana2022ultra,bose2024anneal}


\end{tabular}
\end{ruledtabular}
\end{table*}%

\section{Conclusion}

This work presents the design and experimental demonstration of the heterogeneous photonic integrated a-SiC/SiN platform. The proposed solution combines the merits of both a-SiC and SiN photonics, while circumventing their respective shortcomings. With the a-SiC/SiN platform, we demonstrate a feasible path to leveraging ultra-low optical loss and wide transparent window of SiN photonics, meanwhile featuring up to 4,444 times higher integration density, and 27 times higher thermo-optic tuning efficiency in terms of per unit waveguide length, enabled by a-SiC photonics. The most important element of this heterogeneous platform, a-SiC/SiN interconnection coupler, has been demonstrated to enable efficient interconnections between a-SiC and SiN photonics, with a coupling loss of 0.32$\pm$0.10 dB. Both grating-coupling and side-coupling can be achieved on the same chip, which provides more fiber-to-chip interfacing freedom for distinct requirements from versatile applications. Furthermore, the platform combines two materials that possess high Kerr nonlinear coefficients, which provides the potential of efficient four wave mixing and the realization of on-chip single photon sources. The realization of this heterogeneous a-SiC/SiN platform stands as a compelling candidate towards more advanced applications in integrated photonics, quantum photonics, and beyond.

\section*{Supplementary Material}
Supporting information can be found in Supplementary Material.

\begin{acknowledgments}
The authors acknowledge the help and valuable advices from Thim Zuidwijk, Charles de Boer, and Hozanna Miro. Z.L. acknowledges the China Scholarship Council (CSC, 202206460012). N.S. and I. E. Z. acknowledge the funding from the NWO OTP COMB-O project (project 18757). I. E. Z. acknowledges funding from the European Union’s Horizon Europe research, and the innovation programme under grant agreement No. 101098717 (RESPITE project) and No.101099291 (fastMOT project).
\end{acknowledgments}

\section*{Author declarations}
\subsection*{Conflict of Interest}
The authors have no conflicts to disclose.

\section*{Data Availability}

The data that support the findings of this study are available from the corresponding author upon reasonable request. 

\section*{REFERENCES}
\bibliographystyle{aipnum4-1}
\bibliography{aipsamp}

\end{document}